\documentstyle[aps,eqsecnum,prbbib]{revtex}
\begin{document}
\draft
\title{Nonlinear transverse magnetic moment in anisotropic
superconductors}
\author{Igor \v{Z}uti\'c and Oriol T. Valls}
\address{School of Physics and Astronomy and Minnesota Supercomputer Institute
\\ University of Minnesota \\
Minneapolis, Minnesota 55455-0149}
\date{\today}
\maketitle
\begin{abstract}
We consider the nonlinear transverse magnetic moment that arises in 
the Meissner state of superconductors with a strongly anisotropic order
parameter. We compute this magnetic moment
as a function of applied
field and geometry, assuming d-wave pairing, for realistic 
samples, finite in all three dimensions, of high temperature
superconducting materials. Return
currents, shape effects, and the anisotropy of the penetration depth
tensor are all included. We
numerically solve the nonlinear Maxwell-London equations
for a finite system. Results are discussed in terms of the relevant
parameters. The effect, which is a probe of the
order parameter symmetry in the bulk, not just the surface,
of the sample
should be readily measurable
if pairing is in a d-wave state. Failure to observe it would set a
lower bound to the s-wave component.
\end{abstract}

\section{Introduction}
The question of the pairing state of high temperature superconductors (HTSC's)
continues to baffle researchers in the field. A recent review \cite{agl} of
many of the experimental results closes after over eighty pages  with the
tentative conclusion that the only state compatible with all experiments
would have to exhibit two separate transitions, in contrast with the
single transition invariably observed in HTSC's.
Worse, even different 
experiments performed on the very same single crystal sample 
seem to lead to
contradictory conclusions.\cite{buanyang} This situation makes it
particularly desirable to find good probes that may as unambiguously as
possible discern the structure of the order parameter (OP) function.

One such probe is afforded by the nonlinear Maxwell-London electrodynamics
of exotic pairing states in superconductors in the Meissner state. 
It was pointed out four years
ago\cite{ysprl} in the context of a two-dimensional model
(HTSC's are nearly all highly anisotropic, layered materials)
that the presence of order parameter lines of nodes on a cylindrical Fermi
surface
would lead to observable nonlinear effects in the electrodynamic properties.
These effects were shown to be\cite{ysprl,sv,ys}
potentially
capable of yielding a signature for the structure of the 
order parameter function, and therefore for the nodal structure of the gap
itself. The most obvious of them is the presence of a transverse 
magnetic moment ${\bf m}_\perp$ perpendicular to the direction of an applied
magnetic field ${\bf H}$ lying in the $x-y$ ($a-b$)
plane (the $z$ axis is 
taken to be
along the $c$ crystallographic direction,
perpendicular to the stack of planes in the lattice structure).
This transverse magnetic moment has an angular dependence, as  the sample is
rotated about the $z$ axis, reflecting directly the periodicity of the 
energy spectrum, that is, of the square of the order parameter, on the
azimuthal angle $\phi$. This signal would be particularly prominent if the
order parameter has nodes or very deep minima as a function of angle. 
The transverse moment can be measured directly or inferred from the torque
it produces.

Electrodynamic effects are of particular interest because they probe the
superconducting properties over a scale given by the field penetration depth,
while many other experimental methods, such as tunneling and Josephson
junctions, probe only over the scale of the correlation length, which
in HTSC's is over two orders of magnitude
smaller. Thus, electrodynamics probes the bulk properties of the
superconducting material, not merely the surface. 
Some of the apparent conflicts among
experimental results may be resolved\cite{bahcall}
if the order parameter symmetries
in the bulk and at the surface differ.

For these reasons, investigations of the transverse
magnetic moment effect have continued.
For the model of a ``slab'' shaped sample, infinite in the $x-y$ directions
and of fixed thickness, and a cylindrical 
Fermi surface, the equations
involved in the nonlinear Maxwell-London electrodynamics have been solved
numerically\cite{sv} for the experimentally relevant ranges of applied 
magnetic field
and temperature. At $T=0$, for the same geometry and assumptions, a
perturbative, but accurate, analytic solution
has been obtained\cite{ys}. In Ref.\onlinecite{sv} results were obtained for
both a pure d-wave state and a mixture of $s$ and $d$ waves.
The amplitude of the
transverse magnetization at low temperatures was found to
increase\cite{sv} as approximately the
second power of the applied field, and, since one is dealing with an
effect arising from a finite penetration depth, as the surface area
of the sample. The results\cite{sv,ys} indicated that, for a pure
d-wave state, ${\bf m_\perp}$ should be detectable by a SQUID 
magnetometer in single crystal samples,
even though the applied field is limited,
by the need to keep the
sample in the Meissner regime, to the field of first flux penetration. 

The results of only one experiment seeking to find the transverse
magnetization\cite{confusion}
effect have been published\cite{buan}. No evidence of a transverse
magnetic moment having the proper periodicity was obtained, but the nature of
the data and of its analysis was such that the absence of evidence could 
not be conclusively taken to be evidence of absence.  The
resolution of the measurements was, over most of the field range studied,
comparable to the signal measured. At the higher fields used the
signal was below, but only by a factor of two,
the theoretically expected results for a pure d-wave state, based on
the infinite slab assumptions\cite{sv} discussed in the above paragraph,
scaled to the finite size of the sample. Although a subsequent
analysis\cite{buanyang} of the field dependence of the experimental signal
\cite{buan} was also supportive of the inference that no nodes were
observed, a completely definitive statement about the existence of
nodes could not be made because 
of theoretical and experimental uncertainties.

Although this paper deals with the theoretical questions
it is useful to briefly mention here the main difficulty associated with the
experiments: except in the ideal world of theorists
where all samples, even when finite, are spheres or other highly
symmetric bodies, there {\it always is} in any superconductor, a transverse
magnetic moment because of demagnetization effects\cite{orlando}. For the
rectangular samples used in the experiment\cite{buan} these effects are
much larger than the nonlinear contribution. Of course, their main
periodicity, $\pi$, is twice that
associated  ($\pi/2$) with the d-wave nonlinear effects. Fourier analysis
of the data, as done in Ref.\onlinecite{buan} does filter out the main
spurious demagnetization signal, but its $\pi/2$ harmonic still will be
confounded with the nonlinear signal and, although it can partly be sorted
out\cite{buanyang} because of its different, linear,  magnetic 
field dependence, the
noise introduced in the original signal by the Fourier transforms and
subtractions takes a heavy toll. The work in this paper is motivated in large
part by experiments now being planned\cite{amgpc} on
samples having approximately a flat disk shape that minimizes the 
demagnetization effect for an applied field lying in the $a-b$ plane .
It seems to us that reaching conclusions from experimental data
taken on samples lacking rotational symmetry would be difficult.

The theoretical uncertainties are related to the use in the work
discussed above of two approximations: first, the use of an infinite slab
geometry neglects the contribution to ${\bf m}_\bot$
of return currents flowing along
the sides of the sample, parallel to the z-axis. 
Furthermore, the penetration depth for those currents,
$\lambda_c$ is much larger than that in the a-b plane, $\lambda_{ab}$.
Since ${\bf m}_\perp$ increases with penetration depth, the
question of how the
larger $\lambda_c$ must be included in the analysis arises.
The effect of $\lambda_c$ was included in the analysis
of the data\cite{buan}
but in a purely heuristic way, which assumed that its influence was rather
strong. Although this assumption was relaxed in a subsequent
reanalysis\cite{buanyang}, the whole question remains a major obstacle
in reaching conclusions from experimental evidence.
A finite $\lambda_c$ implies also abandoning the notion of a 
cylindrical Fermi surface. Further, the very neglect of finite size effects is
suspect: they have been shown\cite{lin} to affect the analysis of penetration
depth experiments in rather large samples.
The uncertainties related to these complications
in analyzing experimental data in the context of the geometric
approximations of Ref.\onlinecite{sv} are about as large as the experimental
uncertainties and it might be argued that, in combination, are
about as large as the
discrepancy between theory and experiment.

In this paper we therefore undertake the examination of Maxwell-London
electrodynamics, including nonlinear contributions, for finite
 samples. We
include in the study the effects of $\lambda_c$,
which affect the electromagnetic response even at linear
order in the field,
and we specifically consider geometries relevant to the experimental
systems currently been studied. Because of these realistic assumptions, not
only is an analytic solution unattainable even at zero temperature, but the
numerical work involved is quite considerable. Our results confirm that
a transverse magnetic moment  should be observable in obtainable
single crystal samples, if the bulk pairing is in a d-wave state.
The effect discussed in this paper is a fingerprint for the
existence of an OP with d-wave symmetry. Its experimental observation
in suitable samples, would constitute very hard to refute evidence for
d-wave pairing in the bulk. Failure to observe this effect would at the
very least put a lower bound on the existence of an s-wave component.

In the next Section we present the equations solved and we discuss the 
methods we use. In Section III, we present our results and predictions for
experimental outcomes, assuming a d-wave order parameter is present. Our
conclusions are given in the last Section.

\section{Methods}
\label{sec:methods}
\subsection{Maxwell-London electrodynamics}
The equations of superconducting electrodynamics are in textbooks
\cite{orlando,degennes} and need no rederivation. We will merely 
introduce here our notation, and briefly recall a couple of easy to
overlook
points. For the linear case, a particularly clear discussion of the
complications that occur when the penetration depth tensor is anisotropic
is in Chapter 3 of Ref.(\onlinecite{orlando}).

Outside the sample the current is ${\bf j}=0$ and therefore in the
steady state the field ${\bf H}$ satisfies the Maxwell equations 
$\nabla \cdot {\bf H}=0$ and $\nabla \times {\bf H}=0$. One can therefore
write ${\bf H}$ in terms of a scalar potential $\Phi$ satisfying the Laplace
equation:
\begin{mathletters}
\label{out}
\begin{equation}
{\bf H}=-\nabla \Phi  \label{equationa}
\end{equation}
\begin{equation}
\nabla^2 \Phi =0 \label{equationb}
\end{equation}
\end{mathletters}
In the interior of the sample one must consider three fields: ${\bf H}$,
${\bf j}$ and the ``superfluid
velocity field'' ${\bf v}$ defined as:
\begin{equation}
{\bf v}=\frac{\nabla \chi}{2} + \frac{e}{c} {\bf A},
\label{vdef}
\end{equation}
where $\chi$ is the phase of the superconducting order parameter, ${\bf A}$
the vector potential, and $e$ the
proton charge. We set $\hbar=k_B=1$.
Eq.(\ref{vdef}) is equivalent to Eq. (5.99) in 
Ref.\onlinecite{orlando}. The field ${\bf v}$ has dimensions of momentum, not
velocity, but the factors associated with the effective mass and its 
anisotropies are more conveniently dealt with by placing them elsewhere.
The three equations that one requires for these three fields are first, the
second London equation obtained by taking the curl of (\ref{vdef}):
\begin{equation}
\nabla \times {\bf v}=\frac{e}{c}{\bf H} 
\label{londoneq}
\end{equation}
second, Amp\`ere's law:
\begin{equation}
\nabla\times {\bf H}=\frac {4 \pi}{c}{\bf j}
\label{ampere}
\end{equation}
and finally, a constitutive equation relating ${\bf j}$ and ${\bf v}$ which
is  discussed in general in the next subsection. For the usual linear
case this relation is of course simply:
\begin{equation}
{\bf j}=-e\tilde{\rho}{\bf v}
\label{linear}
\end{equation}
where $\tilde{\rho}$ is the superfluid density tensor. The nonlinear
contribution is discussed below (see Eq. (\ref{jtzero})).
$\tilde{\rho}$ is related to
the penetration depth tensor $\tilde{\Lambda}$, whose components in the
diagonal representation are the square of the London penetration
depths, by the relation:
\begin{equation}
\tilde{\rho}=\frac{c^2}{4 \pi e^2} \tilde{\Lambda}^{-1}
\label{rholam}
\end{equation}
When these tensors are
proportional to the identity, then one can combine Eqns.(\ref{londoneq}) and
(\ref{ampere}) and find that any one of the three fields considered satisfies
the vector Helmholtz equation. But this is not true\cite{orlando} when 
$\tilde{\rho}$ is anisotropic. One still has, however, the completely general
equation:
\begin{equation}
\nabla\times\nabla\times{\bf v}=\frac{4 \pi e}{c^2}{\bf j(v)}
\label {maxlon}
\end{equation}
which is valid whatever the relation ${\bf j(v)}$
might be. Eq. (\ref{maxlon}) will be the basic equation we will consider
here.

These equations must be solved with appropriate boundary conditions. These are:
first, at infinity (that is, very far away from the finite 
sample) ${\bf H}$ must reduce to the applied field. Second, deep inside the
sample all fields must vanish. Finally, at the interface ${\bf H}$ must
be continuous\cite{degennes,reitz} and the component of ${\bf j}$ normal to
the interface must vanish.

Once the currents ${\bf j}$ inside the sample are known, the magnetic
moment can be obtained by integration:\cite{jackson}
\begin{equation}
{\bf m}=\frac{1}{2 c}\int d{\bf r} \: {\bf r}\times {\bf j(v(r))}
\label{magmom}
\end{equation}

It is convenient to rewrite this equation in terms of surface integrals.
Using Eq. (\ref{ampere}) and  formulas for vector calculus \cite{jackson2}
we derive:
\begin{eqnarray}
{\bf m} &=\frac{1}{8 \pi}\int_{s} d^2S \: {\bf n} \: ({\bf r}\cdot {\bf H})
+\frac{1}{8 \pi}\int_{s} d^2S \: {\bf r} \: ({\bf H}\cdot {\bf n}) \nonumber \\
        &- \frac{1}{8 \pi}\int_{s} d^2S \: {\bf H} \: ({\bf n}\cdot {\bf r})
         +\frac{c}{8 e\pi}\int_{s} d^2S \: {\bf n} \times {\bf v}
\label{eq:magmoms}
\end{eqnarray}
Integration is performed over the sample surface S,  {\bf r} is the 
position vector for a point on the surface S, and {\bf n} is the 
unit normal pointing outwards.
         
In the linear regime, and for the case which is experimentally relevant
here where one has a small but finite value of the ratio $\lambda/d$
between a typical penetration depth $\lambda$ and a characteristic sample
dimension $d$, the components of ${\bf m}$ can generically
be written in the form $m=m_0(1-\alpha(\lambda/d)+{\it O}(\lambda/d)^2)$.
Examples for values of the
positive constant $\alpha$ are given in textbooks\cite{fetwal,london}.
One has, then, a reduction in the magnetic moment due to current
penetration in the material. 
The nonlinear effects may be conveniently viewed for  our purposes as
anisotropic, field-dependent corrections to the values of the $\alpha$
coefficients.

\subsection{The supercurrent}
To obtain the supercurrent response 
one has to specify the full constitutive relation 
${\bf j(v)}$ as discussed above.
The linear contribution has been given in Eqns. (\ref{linear}) and 
(\ref{rholam}).
The anisotropy in $\tilde{\Lambda}$ is related to that of the Fermi surface and
can be related to the effective mass tensor\cite{orlando}.
The anisotropy
of the order parameter considered in this
work would not, if the Fermi surface
were isotropic, lead  to
any anisotropies in the penetration
depth\cite{sv}, at linear order in the magnetic field. Thus we have:
\begin{equation}
\tilde{\Lambda}=\frac{c^2 \tilde{m}}{4 \pi n e^2}
\label{lamm}
\end{equation}
where $\tilde{m}$ is the effective mass tensor and $n$ the carrier
density. 

The nonlinear corrections to ${\bf j(v)}$ were first discussed by
Bardeen\cite{bardeen} in the context of an s-wave superconductor.
In that case,
at $T=0$, the current-velocity
relation is linear for velocities less than a critical 
velocity $v_c$ defined below. Nonlinear corrections to the 
current-velocity relation arise from the thermal population of quasiparticles. 
These corrections are small\cite{ys} (cubic in the ratio $v/v_c$) and 
in the regime of larger flow velocities vortex nucleation may occur before 
these effects become important\cite{degennes}. On the other hand, 
in unconventional superconductors, particularly those
with nodes in the OP, nonlinear 
corrections are significantly larger and qualitatively different than 
in a superconductor with an s-wave excitation gap. The region on the 
Fermi surface (FS) close 
to the nodes in the gap provides enhanced nonlinear response due to 
the higher population of quasiparticles. There the energy required to break 
electron pairs will be reduced and will vanish along the direction of the
lines of nodes in the gap. Thus nonlinear corrections in the presence of 
nodes  in the OP are nonvanishing even at $T=0$ and the nonlinear behavior 
exhibits anisotropy with respect to the relative angle between applied
magnetic field
and the lines of nodes of the gap function. 

For the purpose of computing ${\bf m_\perp}$
for a crystallographycally strongly
anisotropic HTSC,
in the geometry considered here,
we are concerned with the angular dependence 
of the OP in the plane of the applied field, the $a-b$ plane. 
In most of the numerical calculations performed in this  work, we consider an
order parameter of the pure d-wave form:
\begin{equation}
\Delta=\Delta_0 \sin(2 \phi)
\label{op}
\end{equation}
where $\phi$ is the azimuthal angle referred to a node and $\Delta_0$
the amplitude.  
The periodicity
of ${\bf m_\perp}$, which equals that of the energy, is therefore $\pi/2$.
It would be possible, as we shall see, to generalize
the calculations to other $\phi$ dependence for the OP if necessary.
Possible dependence of $\Delta$ on the $c$ direction is very
uncertain because of multilayering effects\cite{agl}, 
but we believe it is unlikely to be important. This
follows from the above physical considerations, and from the brief discussion
of results for alternative forms in the next Section and Appendix A.
On the other hand, we must properly take into account the strong anisotropy
of $\tilde{\Lambda}$ in the $c$ direction, which is related to that of
the Fermi surface.

The assumption of isotropy in the $a-b$ plane deserves further discussion,
since anisotropies in the in-plane penetration depth in HTSC's are
known to exist.\cite{kim}
Anisotropy in $\lambda_{ab}$ will lead to a contribution
to ${\bf m_\perp}$ of periodicity $\pi$ (instead of $\pi/2$ for the nonlinear
signal), and (see the paragraph below (\ref{eq:magmoms})) down by a factor
of $\sim \lambda/d$ (the ratio of a typical penetration depth to 
a typical sample dimension) from the longitudinal, linear magnetic moment.
This contribution will be linear, not quadratic, in
the field. Hence, from an experimental point of view, it is in
effect a small correction to the effective demagnetization factor 
which is, although to a small extent, present even in macroscopically
symmetric samples. Thus, Fourier analysis of the experimental signal
and examination of the field dependence of the harmonics would
separate this from the nonlinear effect. As for the theoretical results
discussed here it is sufficient to interpret our symbol
$\lambda_{ab}$ for the  $a-b$ plane penetration
depth ($\lambda_c$ is that along the c-direction)
as the geometric mean of $\lambda_a$
and $\lambda_b$. Finally, if because of orthorhombic anisotropy the
nodes are not precisely at $\pi/2$ angles, this can be accounted for
by adding a small constant\cite{agl} to (\ref{op}) and therefore the
$\pi/2$ Fourier component of the response would be little affected.

For an order parameter with nodes, it is known\cite{sv,ys} that the
details of the FS shape are not important
for the nonlinear properties we deal with here: only the
nodes and their symmetry matter. Thus, our chief concern about the FS is
to describe the anisotropy in $\tilde{\Lambda}$. For this purpose, 
we have used in our calculations an ellipsoidal (of revolution)
Fermi surface characterized by effective masses $m_{ab}$ (in the $a-b$ plane)
and $m_c$, as in Eq. (\ref{ef}).  
We introduce the ratio $\delta\equiv (m_c/m_{ab})$.
We define a speed $v_f$ in terms of the Fermi
energy $\epsilon_f$ as $v_{f}^{2}=2 \epsilon_f /m_{ab}$ which facilitates
comparison with two-dimensional results.

The general expression 
\cite{sv,ys} for ${\bf j(v)}$, including nonlinear terms, can be written as:
\begin{equation}
j=-eN_f\int d^2s \: n(s) {\bf v}_f [({\bf v}_f \cdot {\bf v})
  +2\int^{\infty}_0 d\xi \: f(E(\xi)+{\bf v}_f\cdot {\bf v})]
\label{jtzero}
\end{equation}
where  $N_f$ is the total density of states at the Fermi Level,
$n(s)$ is the density of states at point $s$ at the Fermi surface,
normalized to unity, ${\bf v}_f(s)$ is the $s$-dependent Fermi velocity,
$f$ the Fermi function, and
$E=\sqrt{\xi^2+\left| \Delta(s) \right|^2}$.
In general, the integrals in Eq. (\ref{jtzero}) can be evaluated only
numerically, as was done in 2-d in Ref.\onlinecite{sv}. 
However it was shown
there that at the temperatures (about two Kelvin)
where the experiments are performed, one is very near the zero
temperature limit. Therefore  we confine ourselves in this work to this
limit, where, within the assumptions discussed above,
it is possible to derive an analytic
expression for ${\bf j(v)}$, valid in the limit where $\delta$ is large.
Having an analytic expression for ${\bf j(v)}$ makes the subsequent 
substantial numerical work easier.
In a magnetic field (hence ${\bf v}\neq 0$) even at $T=0$ there 
exists a region with $\left|\Delta(s) \right|+{\bf v}_f \cdot {\bf v}<0$ 
where it is  possible to have a quasiparticle population. One can then
perform, as in Ref.\onlinecite{ys}
an approximate but accurate integration of the general formula 
Eq. (\ref{jtzero}),
to obtain 
the nonlinear corrections at zero temperature (details are given in 
Appendix A).
The resulting ${\bf j(v)}$ relation is valid under the
assumptions for the FS and the OP discussed above, and for any
three dimensional strongly anisotropic
superconductor (independent of sample geometry). Introducing
cartesian axes x'-y' fixed to the crystal along the nodal directions
the expression we obtain is:
\begin{mathletters}
\label{jnl}
\begin{equation}
j_{x',y'}  =-e\rho_{ab} v_{x',y'} (1-\frac{9\pi}{64 v_c}\left|v_{x',y'}
\right|) 
\end{equation}
\begin{equation}
j_z =-e\rho_{c} v_{z} (1-\frac{3\pi}{32 v_c}\frac{v_{x'}^2+v_{y'}^2}
{\left|v_{x'}\right|+\left|v_{y'}\right|})
\end{equation}
\end{mathletters} 
where $v_c\equiv \Delta_0/ v_f$. The
components of the superfluid density tensor are
$\rho_{ab}=\frac{1}{3}N_f v_f^2$, $\:$ $\rho_{c}=\rho_{ab}/\delta$,
in the $a-b$ plane and along the 
$c$ crystallographic axis respectively. 
Even for YBCO reported\cite{erange} measured values for 
$\delta=m_c/m_{ab}=(\lambda_c/\lambda_{ab})^2$
are quite large, between 15 and 50. Omitted terms 
in the above  relation are  those involving higher powers of 
$v/v_c \ll$ 1 (for the regime of experimental interest in the Meissner 
state) or supressed by the small parameter $\delta^{-1}$.

\subsection{Geometry}
In this subsection we discuss the sample geometry considered in the calculation
of the supercurrent response in the Meissner state. The dependence
of measured 
 quantities such as the local magnetic field, actual current distribution
and magnetic moment on sample shape is one of the issues of particular 
interest in this work, as explained in the Introduction.

The experimentally relevant regime is that of small but 
finite ratio of penetration depths to typical sample dimension. 
As mentioned below Eqn.(\ref{eq:magmoms}), the geometric dependence of
the effect under investigation in this regime\cite{film} is proportional to
these
ratios. Thus, the coefficient of this finite size term must be accurately
calculated. A ``locally flat'' approximation (assuming a purely exponential
decay for the fields away from the surface) is not sufficient for this
purpose, as one can check even in the linear case from the exact solution
for a sphere\cite{reitz}.  
It is therefore necessary to solve the complete boundary value
problem,that is, to find the solution to the equations
(\ref{out}) and (\ref{maxlon}) outside and inside of the sample. 
This is in principle a numerically difficult undertaking, since {\it a priori}
it involves solving a system of partial differential equations with nonlinear
terms in the entire three dimensional space, but with currents being
confined to a very small region where great precision is required. 
Although sophisticated variable grid methods could perhaps be found to 
take care of these complications, the relatively symmetric
shape of the required experimental samples allows for a simpler
approach.

Single crystals of HTSC materials are typically flat, much thinner in the
c-direction than in the  $a-b$ plane. The magnetic field, we recall, is
applied parallel to this plane along the $x$ direction.
The geometry and coordinate
system we consider are sketched in Fig. \ref{fig1}.
Because of the complications associated with demagnetization factors,
as described in the Introduction, experiments  must be performed
on single crystals of a highly symmetric shape. This is
achieved by  laser cutting, or shaving\cite{bgg} the crystals so that
their cross section
in the direction perpendicular to the $z$ axis
is circular. Their shape is, therefore, roughly a disk, thinning towards
the edges because of mechanical disintegration associated with the cutting.
They have smooth edges and
can therefore be described as ellipsoids of revolution. 
This geometry has also a considerable computational advantage: the
form of the solution outside the
sample in the limit $\tilde{\Lambda}=0$ is known exactly
since the potential $\Phi$ (see (\ref{out}))
satisfies trivial Neumann boundary conditions at the interface and can be found
by electrostatic analogy.
The solution contains a single parameter which is simply related to $m_0$,
the value of the 
longitudinal magnetic moment in the zero penetration limit.
When the penetration depth
is finite, the longitudinal moment does change, but its correct
value can be determined from the boundary conditions and the solution
inside through an iteration process as described in the next Section.
But, as important as these simplifications are, they only go so far:
the fundamental equation (\ref{maxlon}) is not separable in spheroidal
coordinates.

We therefore consider a supercondutor in the shape of a flat ellipsoid 
of revolution (an oblate spheroid) with the axis of revolution along 
the z-axis. Its major and minor semiaxes are denoted by $A$ and $C$ 
respectively,  and we have $A > C$ for
actual samples. We take
(see Fig. \ref{fig1}) a coordinate system fixed to the direction
of the magnetic field, with its z-axis
parallel to the $c$ crystalographic direction of the superconductor (and
parallel to the $C$ semiaxis of an ellipsoid). 
The field is applied along the x-axis, and we picture the experiment
as being performed by rotating the crystal about the z-axis. 
As the crystal is rotated the axes $x-y$ remain fixed in space, and 
should not be confused with the
previously introduced $x'-y'$ coordinates, affixed to the
crystal structure.

Even with the nonseparability of the equations, it is still convenient from the
point of view of fulfilling the interface boundary conditions,
to introduce oblate spheroidal coordinates $\alpha, \beta, \varphi$. 
In the definition we use\cite{arfken}, they are related to cartesian   
coordinates by the transformation:
\begin{mathletters}
\label{eq:coor}
\begin{equation}
	x = f \cosh \alpha \sin \beta \cos \varphi \\
\label{coora}
\end{equation}
\begin{equation}
        y = f \cosh \alpha \sin \beta \sin \varphi \\
\label{coorb}
\end{equation}
\begin{equation}
        z = f \sinh \alpha \cos \beta
\label{coorc}
\end{equation}
\end{mathletters}
where $0\leq \alpha<\infty, 
0\leq\beta\leq\pi, 0\leq\varphi\leq 2\pi$,
and $f$ is a focal length scale factor. The surface of an ellipsoid corresponds
to a given value of $\alpha$.

For an ellipsoid of revolution (about its $z$ axis) in the presence of 
uniform magnetic field $\bf{ H}_a$ applied along the direction that we
denote as the $x$ axis  the 
magnetic potential in the outside region for $\lambda=0$ 
has the form \cite{russian}
$\Phi=-H_a x(1+g(\sinh\alpha))$, where the
first term is the potential for the uniform applied field ${\bf H}_a$ 
and the gradient of the second term is the field generated by 
the superconducting ellipsoid. Writing out in detail the function
$g$ one has:
\begin{equation}
\Phi=-H_a f P^1_1(i\sinh\alpha) P^1_1(\cos\beta)\cos\varphi+
A_1 f Q^1_1(i\sinh\alpha) P^1_1(\cos\beta) \cos\varphi
\label{eq:phiout} 
\end{equation} 
where $P^{1}_{1}$ and $Q^{1}_{1}$ are associated Legendre functions of 
the first and second kind respectively. 
The parameter $A_1$
is determined from the boundary conditions. 
It is proportional to $m_{\|}$, the longitudinal magnetic
moment:
\begin{equation}
m_{\|}=\frac{2}{3}f^3A_1
\label{longmag}
\end{equation}
Its value  at $\lambda=0$ is:
\begin{equation}
A_1(0)=\frac{- H_a\sinh\alpha_0}{1+1/\cosh^2\alpha_0-\sinh\alpha_0
\arctan(1/\sinh\alpha_0)}
\label{a1}
\end{equation}
where $\alpha=\alpha_0$ is the value at the surface of the ellipsoid.
Eqns.(\ref{a1}) and (\ref{longmag}) can be combined to yield the usual
expression for $m_0$, the zero penetration depth, purely longitudinal,
magnetic  moment, in terms of the ellipsoidal demagnetization
factors.\cite{ll}
At finite $\tilde{\Lambda}$ the value for 
$A_1$ is obtained from the iterative procedure  
discussed in the next subsection. 
In the spherical limit $\sinh\alpha\rightarrow\infty$ we also recover the 
standard result\cite{reitz,london}.

\section{Results}
\subsection{Numerical procedure}              
 
Let us discuss now the iterative
procedure we use to solve (\ref{maxlon}), while avoiding to have to numerically
solve the field equations in all space. For clarity, let us focus first on the
case where the nonlinear effects are neglected but the penetration depths are
finite. This has two effects on the solution outside: first, the value of
$A_1$ (or of $m_{\|}$) deviates from the value given by (\ref{a1}). Secondly,
the field acquires higher order multipoles, i.e. the field potential 
acquires additional terms:
\begin{equation}
\Phi = \Phi_a +A_1 f Q^1_1(i\sinh\alpha) P^1_1(\cos\beta) \cos\varphi
+\sum_{n\geq 1}A_{n} f Q^1_n(i\sinh\alpha) P^1_n(\cos\beta) \cos\varphi 
\label{figen}
\end{equation}
where $\Phi_a$ is the potential corresponding to the applied field (the
first term on the right side of (\ref{eq:phiout})) and the gradient of the
rest is the field created outside the sample by the current distribution in
it.
The terms with $n>1$ vanish at $\tilde{\Lambda}=0$ (and also in the spherical
case if $\tilde{\Lambda}$ is
isotropic.) The fields generated by $A_1$ are not purely dipolar, since they
have ellipsoidal symmetry. However, 
the dipole moment is determined
by this parameter only. 
At the first step of the iteration we solve the linear version of
(\ref{maxlon}) for the current field, assuming that the outside field is
given by its $\lambda=0$ limit, i.e. the gradient of (\ref{eq:phiout})
with (\ref{a1}). Parameter counting shows that in order to do so we must
give up one boundary condition, and accordingly we temporarily sacrifice
the continuity of the ``radial'' field ($\hat{\alpha}$ component). From the 
resulting current distribution we compute ${\bf m}$ through 
(\ref{eq:magmoms}). This is of course not the same as the input moment.
We then replace this obtained value in the external potential (through 
(\ref{longmag})) and repeat the procedure\cite{flat}
until the moment generated by
the computed currents equals the input value.
The iteration is considerably simplified by observing that
several terms  in (\ref{eq:magmoms}) vanish explicitly when $\tilde{\Lambda}=0$
and by making use of the fact that the penetration
depths are small\cite{zv} compared to the sample dimensions.
Once the iteration is concluded, one finds that
$H_\alpha$ is continuous except in a very narrow band near the equator
corresponding to a symmetry higher than dipolar. This can then be eliminated
by adding small values for
higher order $A_n$'s to the expansion (\ref{figen}), but symmetry
considerations and examination of Eq. (\ref{eq:magmoms}) show that these
additions do not affect the already determined value of the sample magnetic
moment.  
It is easy and very instructive
to verify analytically that this procedure recovers the known
result\cite{reitz} for the $\lambda$ dependent magnetic moment of a sphere
with an isotropic $\tilde{\Lambda}$. 

The same procedure is used with the nonlinear terms, with only two important
differences: first, to the field outside one must add a transverse dipole,
i.e. a contribution of the form of the last term in (\ref{eq:phiout}) rotated
$90^\circ$:
\begin{equation}
\Phi_{\bot}=A_{1\bot} f Q^1_1(i\sinh\alpha) P^1_1(\cos\beta) \sin\varphi
\label{eq:phiout2} 
\end{equation}
where $\frac{2}{3}f^3 A_{1\bot}=m_\bot$. Of course this term does not
exist when the penetration depths vanish, since the nonlinear effects are
absent unless the field can penetrate the sample.
In principle one should also
consider higher order multipole terms, as in (\ref{figen}) but we have found
that any such terms are below the precision level of our numerical results.
Both components of the moment
${\bf m}$ are determined through the vector relation (\ref{eq:magmoms}).
The other important difference is the obvious one of using (\ref{jnl}) for the
${\bf j(v)}$ relation. There are also practical differences, however,
since in the linear equations the variable $\varphi$ can be separated out
while in the full nonlinear case all three coordinates are coupled.

In the actual solution of the equations we use a relaxation method. We
proceed in two steps: first we solve the linear problem, which involves
only two variables, since then the $\varphi$ dependence of all quantities
can be determined analytically. The iterated solution for that
problem is then used as the initial guess in the full three-variable
nonlinear problem.

We discretize the differential equations expressed in ellipsoidal
(oblate spheroidal) coordinates, on a three dimenssional grid. 
An obvious advantage of  the
ellipsoidal grid is that it simplifies consideration of boundary 
conditions at the surface of the ellipsoid, given by equation 
(\ref{londoneq}). Numerically, it is more intricate to consider boundary 
conditions that involve derivatives, and accuracy is increased if the
grid points are also boundary points. 

The discretization procedure we have used has an estimated error quadratic in
the spacing between the grid points. All quantities involved in
equations (\ref{maxlon}) have definite parity with respect to the exchange $z$
$\rightarrow$$-z$ and it is sufficient to solve them only for one half of 
the sample. 
Accuracy is predominantly governed
by the spacing between grid points along the $\hat{\alpha}$ 
direction. In the actual numerical solution (for half the sample), 
we consider ellipsoids 
of different shapes (different $C/A$) and sizes (different $A$). 
It is appropriate to increase $n_\alpha$ proportionally to $C/A$.
Denoting by $n_{\alpha}$ the  number of grid points along 
the $\hat{\alpha}$ direction, 
 the smallest $n_{\alpha}$ used was 100, and the largest
800, spaced in the region of nonvanishing currents. The number of grid 
points along the  $\hat{\beta}$ and 
$\hat{\varphi}$ directions was $n_{\beta}=50$ and $n_{\varphi}=30$ 
respectively.  Increasing these numbers by a factor of two gave only effects 
below the numerical accuracy attained, which is about two significant
figures, an error much smaller that the uncertainty arising from 
the imprecise knowledge of the experimental values of the input
parameters.

As one of the checks on 
the accuracy of the algorithm we use, it is instructive to 
consider the case of an isotropic spherical sample, (both the sample and
the Fermi surface are spheres, $\delta=1$) in the linear regime,
where the analytical solution is known. To ensure that we tested the same
algorithm, we treated the sphere as the 
the spherical limit of ellipsoidal coordinates, which 
corresponds to $\sinh \alpha$ $\rightarrow \infty$. 
We used  $\sinh \alpha$ =1000, equivalent to 
eccentricity  $e=\sqrt{1-(C/A)^2}$=0.001 with the grid given by 
$n_{\alpha}=200$, $n_{\beta}=50$ and $n_{\varphi}=30$ grid points. We 
solved equations (\ref{maxlon}) in a spherical shell of thickness 10$\lambda$ 
and studied both the magnetic moment and the current distributions.
The accuracy for the current on any grid point corresponded to 
four significant figures for the region where currents are important.
The magnetic moment calculated both from 
(\ref{magmom}) and from the surface integrals
(\ref{eq:magmoms}) agrees with the exact result, including the correct
finite penetration depth correction. 

We have also checked that our results for the longitudinal magnetic moment
extrapolated to the zero penetration depth limit agree with the known
analytic result for ellipsoids. We have also
verified that the magnetic moment calculated from 
(\ref{magmom}) agrees from that found from the surface integrals
(\ref{eq:magmoms}). The latter procedure is, however, much more convenient.

\subsection{Numerical results and discussion}
In performing the calculations and describing the results, it is
convenient to introduce dimensionless quantities. Because of the shape
of the samples, we use $\lambda_{ab}$
as the unit of length.  We then  define 
the dimensionless fields:
${\bf V}$, ${\bf J}$, and ${\bf h}$:
\begin{equation}
{\bf V}=\frac{\bf v}{v_c},\qquad {\bf h}=\frac{{\bf H}}{H_0},\qquad 
{\bf J}=\frac{c H_0}{4\pi \lambda_{ab}} \, {\bf j},
\label{eq:dimless}
\end{equation}
where we have introduced 
a characteristic magnetic field $H_0$ as: 
\begin{equation}
H_0=\frac {\phi_0}{\pi ^2 \lambda_{ab} \xi_0}
\label{h0}
\end{equation}
where $\phi_0$ is the flux quantum and $\xi_0=v_f/\pi \Delta_0$ is the 
in-plane superconducting coherence length. 
The definition  of
(\ref{h0}) involves precisely the same numerical factors as that
used in Ref.\onlinecite{sv}.
The required equations are easily rewritten in terms of these quantities.
The boundary conditions for the velocity field in (\ref{maxlon}) are now:
\begin{equation}
\nabla\times {\bf V}={\bf h} \mid _{\alpha=\alpha_0} 
\label{eq:lon2}
\end{equation}
where the right hand side is the external dimensionless field at the
surface of the ellipsoid
and from now on the derivatives are with respect to dimensionless length. 
The remaining boundary  condition that there
is no normal component of current at the surface is readily 
obtained from Eqs. (\ref{jnl}). 
The relation between ${\bf j}$ and ${\bf v}$ in
equations (\ref{jnl}) then becomes:
\begin{mathletters}
\label{maxlon2}
\begin{equation}
( \nabla\times\nabla\times {\bf V})_{x',y'} =- V_{x',y'} 
(1-\frac{9\pi}{64}\left|V_{x',y'}
\right|) 
\label{maxlon2a}
\end {equation}
\begin{equation}
(\delta \nabla\times\nabla\times {\bf V})_{z} =
-V_z(1-\frac{3 \pi}{32}\frac{V_{x'}^2+V_{y'}^2}
{\left|V_{x'}\right|+\left|V_{y'}\right|})
\label{maxlon2b}
\end{equation}
\end{mathletters}
where we recall $\delta=(\lambda_c/\lambda_{ab})^2=m_c/m_{ab}$.
Equations (\ref{maxlon2})  are transformed to ellipsoidal coordinates, as
defined above.
Expressions for the superfluid density tensor and  the differential operator 
$\nabla\times\nabla \times {\bf V}$, in ellipsoidal coordinates are 
included in Appendix B. 

In Figures \ref{linearcur}
and \ref{nonlincur} we show some of the results for the currents. These
figures illustrate some of the physics, as well as the quality of the
numerics. In Fig \ref{linearcur}
we show the current $j_z$ going along the $z$ direction in the 
$x-y$ plane as a function of distance from the
surface of the sample starting at the point with
cartesian coordinates $(0,A,0)$. This current
is overwhelmingly determined by the usual linear response to the field.
One can clearly see that its
decay as a function of
depth from the surface
is governed by the $\lambda_c$ penetration depth, and not by
$\lambda_{ab}$. The next Figure illustrates the difference between the
linear and nonlinear components of the current. It shows the current,
again as a function of depth, starting at the center of the top
of the sample, i.e. a point with  cartesian coordinates $(0,0,C)$. 
The component $j_y$, normal to the applied field, is very predominantly
``linear''
and one can see that this time the 
decay is governed by the in-plane penetration depth, as expected
from the geometry. The component $j_x$ along the
applied field, on the other hand,
arises exclusively from nonlinear effects: symmetry considerations
show that it vanishes in the linear limit. Its overall
scale is down by a large factor (basically the ratio of longitudinal
and transverse moments). One sees that even though
the overall decay of $j_x$ is
determined by the scale $\lambda_{ab}$, its behavior is very far from
exponential: it changes sign as a function
of depth. This can be readily understood: at the positions plotted,
close to the center of the 
the top of the ellipsoid,  
$j_x$ is approximately proportional (from Amp\`ere's law)    
to the derivative
with respect to $z$ of the anomalous component of the field, $H_y$. This
field component
nearly vanishes at the surface (it would vanish for a slab) and decreases
exponentially at depths larger than $\lambda_{ab}$. Hence, at this
position in the sample, $H_y$ has an extremum
as a function of depth, and its derivative with respect to
$z$ {\it must} change sign at some point, as we find. This
shows the delicate intricacy of the nonlinear current patterns inside the
material.

Before proceeding with the detailed
discussion of  the dependence of our results on 
the relevant physical quantitites (i.e. size, shape, applied field, and
penetration depths) we illustrate their general scope 
by describing our prediction for the transverse magnetic moment
of a possible HTSC superconducting sample, assumed
to be in a pure d-wave pairing state. This is done in Fig. \ref{predict}.
We show there results for $m_\perp$ as  a function of applied
field. The quantity shown is the maximum value of $m_\perp$ as the crystal
is rotated. It is assumed that the sample is an ellipsoid with $A=2$ mm
and $C=0.1$ mm. Material parameters are taken to be $\xi_0=20\AA$,
$\lambda_{ab}=1800\AA$, and $\lambda_c=9000\AA$. The characteristic
field $H_0$ would be about 5800 gauss. One can see that the
magnetic moments involved are readily accessible to measurement. 
Predictions for samples of other sizes, shapes and material parameters
can be conveniently extracted from the information given below.

We have obtained results for the transverse magnetic
moment for a wide range of the experimentally accessible values of the
appropriate dimensionless parameters, which  as we shall see, can be taken to be
${H_a}/{H_0}$, $\lambda_c/\lambda_{ab}$, ${\lambda_{ab}}/{A}$,
and ${C}/{A}$, the aspect ratio of the ellipsoid of
revolution. We did not consider in our study the ``thin film'' situation
where the sample is so small or so thin that its relevant dimensions
are comparable to or smaller than
the corresponding penetration depth. This case would be
of no interest since the nonlinear effects then are vanishingly small\cite{sv},
and it is excluded from the analysis that follows.

We begin our general discussion of the
results by performing some dimensional analysis.
The quantity $4 \pi m_\bot$ has 
dimensions of magnetic field times volume. The expression 
\begin{equation}
Q=\frac{4 \pi m_{\bot}}{H_a V}
\label{Qeq}
\end{equation}
where $V$ is the sample volume,
is therefore dimensionless. Even though $Q$ is suitable for some 
purposes, it is more convenient to analyze the dependence of the results on
sample size in a different way. The reason is that $4\pi m_\bot$
does not scale 
with the sample volume but\cite{sv} as its surface area. The coefficient
of proportionality  between transverse moment and area
depends on the sample {\it shape} and one can express
this dependence through the aspect
ratio\cite{noecc} ${A}/{C}$. Since the area 
of an ellipsoid of revolution is $\pi A^2$ times a function of ${A}/{C}$,
it is easier for the purpose of giving results in a form more
accessible to experimentalists, to scale explicitly results for ellipsoids
of the same aspect ratio (i.e. the same shape)
by a factor of the ``disk'' area ${\cal S}\equiv \pi A^2$.
The third length that goes into the volume factor in the units
of $m_\bot$ is\cite{sv,ys} a penetration depth. 
Since we are dealing with rather large $A/C$ ratios, with currents
predominantly in the $a-b$ plane, it is advantageous
for our purposes to take this length into accout by writing 
out a factor of $\lambda_{ab}$ explicitly.  Thus we put,
as a first step:
\begin{equation}
4 \pi m_{\bot}(\psi)=
{\gamma}(\frac{A}{C},\frac{\lambda_c}{\lambda_{ab}},H)
{\cal S} \lambda_{ab} f(\psi)
\label{mfirst}
\end{equation}
Here and hereafter we denote the magnitude of the applied field
simply by $H$, rather than $H_a$, as there is no longer a possibility
of confusion.
It is not surprising in view of the above dimensional analysis
that we find 
${\gamma}$ to be a rather weak function of its first two arguments
and independent\cite{scaling} of ${\cal S}$ and of $\lambda_{ab}/A$.
The angular dependence of $m_{\bot}$ is given by the
function $f(\psi$) 
in terms of the angle $\psi$ between
the applied field and a node. We normalize $f(\psi)$ so
that its value is unity at its maximum.

Our results for the angular dependence are shown in Figure \ref{shape}.
The points shown are the values obtained from our numerical calculation.
The error bars indicate the numerical uncertainty. The solid line
represents the  analytic result for the two dimensional
calculations in the slab case\cite{sv,ys}, normalized in the same way. 
We see from the Figure
that the shape
of $f(\psi)$ is, within numerical uncertainty,
the same as for the flat case, where, with the same normalization,
$f(\psi)=3\sqrt{3} \sin\psi\cos\psi(\cos\psi-\sin\psi)$, (for
$0<\psi<\pi/2$).
 This is important, because\cite{sv}
the $\pi/2$ Fourier coefficient of $f(\psi)$, as analytically
calculated from the above expression, 
is unity to three significant figures.  We can therefore identify
here the
coefficient of $f(\psi)$ in (\ref{mfirst}) with the Fourier amplitude
$m_{\bot}$ as introduced earlier in the paper. 
The results shown in this Figure were obtained at $H/H_0=0.1$, $\delta=16$,
and $A/C=19$, but similar results are obtained in all other cases studied.

As can already be seen in Figure \ref{predict}, the
coefficient ${\gamma}$ depends strongly on its third
argument, the applied field. We have studied the field
range $0<H/H_0<0.2$ at $0.05$ intervals. We expect 
that the nonlinear $m_{\bot}$ is proportional to
$H^2$, as in the slab case at zero
temperature. We therefore conclude that a very convenient
way of writing our results for the amplitude is: 
\begin{equation}
4 \pi m_{\bot}= {\cal M}(\frac {A}{C},\frac{\lambda_c}{\lambda_{ab}})
\:\frac{H}{H_0}\:H\:\lambda_{ab}\: {\cal S} \:f(\psi).
\label{mtr2}
\end{equation}
Equation (\ref{mtr2})
implies explicitly that $m_{\bot}/ H$ is a function of field only through
the ratio $H/H_0$.
We have verified that, as expected,
${\cal M}$ is independent of the field. This can be seen in Fig.
\ref{hsquare}, where we plot the quantity
$G\equiv 10^2 4\pi m_\bot/(H\lambda_{ab}{\cal S})$
vs $H$. We see that our results, represented by the symbols with
error bars, are on a straight line, which is indicated by a best fit.
Except for the field, the parameter values are the same as in the
previous Figure.
As in the case of the angular dependence, this qualitative result holds
in all cases studied. The slope of the best fit straight line in plots
such as that in this figure is used to extract ${\cal M}$.

The geometric aspect ratios we have considered in the nonlinear case
range from $A/C=$7 up to 19 (that is from $\sinh\alpha=$0.14433 to
0.05270).  
Comparison
of the results with largest  eccentricity to previous work on the slab case 
illustrates the effect of the return currents. This comparison cannot
be made by taking $C\rightarrow 0$ because we must have
$C>>\lambda_{ab}$. The central portion
of a flat ellipsoid can be identified as a ``slab''. It is simpler to make
the comparison in terms of $Q$ defined in (\ref{Qeq}). It follows from 
our dimensional analysis that: 
\begin{equation}
Q=\tilde{\gamma}\frac{H}{H_0}\frac{\lambda}{2C}f(\psi)
\label{qscale}
\end{equation}
The corresponding result for the slab
is of the same form, and the coefficient may be extracted
from the $T=0$ results of Ref.\onlinecite{ys}. 
When all the relevant factors, such as the different definition of
$H_0$, the normalization of $f(\psi)$, and the setting of the
parameter\cite{ys} $\mu$ (slope of the OP near a node) to $\mu=2$ 
are taken into account,
one finds that the slab value is 
$\tilde{\gamma}=0.056$. We obtain $\tilde{\gamma}=0.092$
for our flattest ellipsoid at $\delta=16$, a number only
weakly dependent on $\delta$. 
Hence, the presence of return currents enhances the nonlinear effects
but, as hinted by the recent reanalysis\cite{buanyang}
of the experimental data,  the enhancement is about three times
smaller, in typical situations,    
than the $\delta$ dependent factor postulated in Ref.
\onlinecite{buan}.

We next show the dependence of our results on the remaining parameters, 
$A/C$ and $\lambda_c/\lambda_{ab}$.
As explained above, our results are most useful to experimentalists if given
in terms of the scaled dimensionless amplitude ${\cal M}$. We therefore
summarize 
the dependence of this quantity on the mentioned length ratios in
Table I. This
table can be used, in conjuction with Eq. (\ref{mtr2}) to
compute the theoretical predictions, when planning experiments
or comparing data and
theory. The rows are the relevant material parameter, given as the
square of the
penetration depth ratio, and the columns are the crystal shape. For
crystals that are not quite ellipsoidal, one should choose a $C/A$
ratio in the table so that the surface to volume ratio of the ellipsoid
agrees with that of the crystal. As the results vary slowly across the
rows and columns of the Table, interpolation and reasonable extrapolation
are obviously feasible. The value of ${\cal M}$ in the Table must be 
multiplied by the square of the applied field and by the cross sectional
area of the sample in the $a-b$ plane, then divided by $H_0$ (which is again
a material-dependent parameter), to obtain the expected value of the
magnetic moment amplitude to be observed. The Table covers anisotropy
parameter $\delta$ values up to 50.

The dependence of ${\cal M}$ on its arguments is weak, less important
than the unavoidable
uncertainty in the values of the experimental input parameters
such as those that go into e.g. the determination of $H_0$. 
The slow dependence of ${\cal M}$
on $\delta$ can be roughly understood as follows:
the portion of the current loop which is along the $c$ direction
does of course contribute to the magnetic moment. However, provided, as in the
Table, that $A>>\lambda_c>\lambda_{ab}$  the exact value of $\lambda_c$
does not matter very much, since (as seen in Fig. \ref{nonlincur})
the decay of the nonlinear current is governed primarily by $\lambda_{ab}$.
The nonlinear moment is then limited by the smaller of the penetration depths.
The slight downtrend with $\delta$ may be due to changes in the complicated
current patterns discussed in connection with Fig. \ref{nonlincur}.
The slight decrease with decreasing eccentricity is likely to be due to the
influence of the $j_z$ component (dominated by linear effects) in a
thicker crystal. The Table indicates that flatter crystals are actually
preferable to harder to obtain thick ones. Again, however, a simple
explanation in intuitive terms is hard to come by, since these results
must be related to the intricacies of the nonlinear current patterns.

Finally we consider the sensitivity of our results to the unknown 
dependence of the order parameter on the crystallographic $c$ direction.
In Appendix A, we have calculated the coefficients of the
nonlinear terms in the current (as in  Eq. 
(\ref{jnl})) for the simplest form of the appropriate\cite{sigrist}
3d OP. We have verified that an increase occurs in the 
quantity ${\cal M}$
but it is negated by a compensating increase in $H_0$, where
then one should replace $\xi_0$ by some average of in and out of plane
correlation lengths. We conclude that the influence of any such dependence
can be neglected as compared with the uncertainty in the material parameters.

\section{conclusions}
The detailed calculations for realistic samples performed here
strongly confirm 
that the transverse magnetization effect 
should be readily observed in available crystals if the order parameter has
nodes of the form expected for d-wave OP
symmetry. From our results, see e.g. Fig. \ref{predict},
it follows that the transverse magnetic moment
generated in a typical sample of about 2 mm diameter and 0.1 mm thickness
(the precise thickness not being very important)
should be $4\pi m_\bot \approx 10^{-6} {\rm gauss\: cm^3}$. 
These numbers are above what can be readily
measured by standard experimental techniques: the rather high
uncertainty levels quoted in past experiments\cite{buan} arise from
the geometrical problems described in the Introduction and also from
easy to overcome uncertainties\cite{anandpc} in the angular positioning
the sample. We have seen that the return currents in the
z-direction produce an enhancement of the effect, although not nearly as
large as that estimated in Ref.\onlinecite{buan}. The effect of the
boundary conditions and the finite size on the result is  somewhat intricate
and, perhaps not surprisingly, we have not been able to find a clear physical
picture in terms of some nonlinear generalization of the demagnetization
factors. Although our results for the influence of sample geometry and
anisotropy in penetration depth were not amenable to  simple
summing up either,
the tabulated values of the quantity ${\cal M}$ should facilitate
experimental design and interpretation.

As briefly indicated in 
Appendix A, (see below Eq.(\ref{phic}))
it is easy at $T=0$ to take into account different
shapes of the d-wave  order parameter function close to the nodes.
Our calculations can also be extended to other symmetries of the OP,
such as an $s+d$ or $s+id$ state, or to take into account
dependence of the OP in the c-direction.
In the worst case, even at
finite temperatures, the functional ${\bf j(v)}$ could be,
numerically
evaluated and used in the calculations, although possibly at considerable
cost in computer time, by means of a lookup 
table and interpolation scheme. 

We have not considered in this work the effect of impurities. This is
unnecessary as it was shown in Ref.\onlinecite{sv} that the impurity 
concentration required to noticeably decrease the nonlinear Meissner effect at
low temperatures is such that it would clearly reduce the transition
temperature of the material. This result will not depend on the geometry of
the sample. On the other hand, we have considered here only the case where
the order parameter has nodes, not just dips. The nonlinear effects are
in fact extremely sensible to the presence of nodes and $m_\bot$ will
decrease substantially, in finite as in infinite samples\cite{sv} if
an s-wave component eliminates the nodes. The magnitude of this decrease
can be gauged from the infinite slab work\cite{sv} since, again, it should
not be excessively sensitive to return currents or sample shape.
Therefore, a negative experimental result could be used to put a 
{\it lower} bound on the amount of s-wave component present.

The methods discussed here can be extended to other sample shapes, to finite
temperatures, and to the computation of other measurable effects that
arise from the same nonlinear phenomena. 

\acknowledgments
We thank A. Bhattacharya, A.M. Goldman, J. Buan
and D. Grupp for many enlightening
conversations concerning the experimental implications of our work, and
B.P. Stojkovi\'c for reading a draft of this work. We also
thank B.Bayman and J. Sauls for discussions. I. \v{Z}. acknowledges
support from the Foster Wheeler and Stanwood Johnston Memorial Fellowships.

\appendix

\section {currents }

We derive here the expression for ${\bf j(v)}$ at $T=0$ starting from equation 
(\ref{jtzero}). Our procedure is a generalization, valid in the
case of interest where the anisotropy $\delta$ is large, of the
two dimesional derivation\cite{ys}. We consider an ellipsoidal 
Fermi surface:
\begin{equation}
\epsilon _f=\frac{k_x^2+k_y^2}{2m_{ab}}+\frac{k_z^2}{2m_c}\equiv
\frac{1}{2} m_{ab} v_f^2                 
\label{ef}
\end{equation}
where we introduce $v_f$ as in Section II. To establish our notation we briefly
review the first, linear term in (\ref{jtzero}):
\begin{equation}
j=-eN_f\int d^2s \: n(s) {\bf v}_f ({\bf v}_f \cdot {\bf v})
\label{firstterm}
\end{equation} 
which is due to the condensate 
contribution to the current. For its evaluation one uses the convenient 
standard textbook method of rescaling $k_z$ by a factor of $\delta^{-1/2}$.
The components of ${\bf v}_f$ are then given by:
\begin{mathletters}
\label{vcomponents}
\begin{equation}
(v_f)_{x'}=v_f \sin \theta \cos \phi \label{vfa}
\end{equation}
\begin{equation}
(v_f)_{y'}=v_f \sin \theta \sin \phi \label{vfb}
\end{equation}
\begin{equation}
(v_f)_{z'}=\frac{v_f}{ \sqrt{\delta}} \cos \theta  \label{vfc}
\end{equation}
\end{mathletters}
where $(\theta,\phi)$, ($\phi$ measured from the x' axis) are the
spherical angles of a vector with components
$k_x, k_y, k_z/\delta^{1/2}$.     
The scalar product in (\ref{firstterm}) can then  be written as:
\begin{equation}
{\bf v}_f \cdot {\bf v}=v_f (v_{x'} \sin \theta \cos \phi
+v_{y'} \sin \theta \sin \phi +\frac{v_{z'}}{ \sqrt{\delta}} \cos \theta)
\label{vfv}
\end{equation}
The integration can be performed by replacing $\int d^2 s \: n(s)$ by
$\int _{\Omega} d \phi \: d \theta \sin \theta /(4 \pi)$ which yields
\begin{equation}
j_{x',y'}=-e \rho_{ab} v_{x',y'} \label{jlina}
\end{equation}
\begin{equation}
j_{z}=-e \rho_{c} v_{z} \label{jlinb}
\end{equation}
With $\rho_{ab}= \frac {1}{3} N_f v_f^2$, $\rho_c=
\rho_{ab}/ \delta$, as quoted in Section II,
(we recall that $N_f$ is the total density of 
states, for both spins). Thus, as is well known,
the effect of the ellipsoidal Fermi surface \cite{muz}
is to rescale $\rho_c$ by a factor of $1/ \delta$ relative to $\rho_{ab}$. 

The nonlinear corrections are contained in the second term of 
(\ref{jtzero}) which is due to the quasiparticle backflow. We denote
this term by ${\bf j}_{qp}$. At T=0, one
can perform the $\xi$ integration: 
\begin{equation}
{\bf j}_{qp}=-2e N_f \int_{\Omega}d \phi \: d \theta \sin \theta /(4 \pi)
\:{\bf v}_f
\: \Theta (-{\bf v}_f \cdot {\bf v} - \left| \Delta (\phi) \right|) 
\sqrt{({\bf v}_f \cdot {\bf v})^2-  \left| \Delta (\phi) \right|^2}
\label{jqp}
\end{equation}  
where $\Delta(\phi)$ is given by (\ref{op}).
In the Meissner regime, 
$\left| v_f v \right|\ll \Delta_0$ 
and the contribution to the quasiparticles arises from 
narrow wedges along the nodal regions, approximately described by the 
azimuthal angle $\phi \lesssim (v/ v_c) \ll 1$.
A superfluid velocity ${\bf v}$ will give rise to backflow 
currents because of its components along nodal regions,
separated by an azimuthal angle $\pi /2$. In two dimensions, ${\bf v}$ can
be\cite{ys} uniquely decomposed into two ``jets'' $v_1 \hat{x}'$ and 
$v_2 \hat{y}'$ along two nodal directions. In the present three-dimensional
case, where the anisotropy $\delta$ is large, one can proceed in a similar
way. We decompose ${\bf v}$ into the sum of two jets 
each directed along a nodal direction and
tilted by the same angle $\omega$ with respect to the z-axis. 
Thus we write
${\bf v}=v_{x'} \hat{x}' +v_{y'} \hat{y}'+v_{z} \hat{z}={\bf v}_1+{\bf v}_2$. 
with:
\begin{equation}
{\bf v}_1=v_{x'} \hat{x}'+\cot \omega |v_{x'}| \hat{z} \label{v1a}
\end{equation}
\begin{equation}
{\bf v}_2=v_{y'} \hat{y}'+\cot \omega |v_{y'}| \hat{z} \label{v1b}
\end{equation}
\begin{equation}
\cot \omega=\frac {v_z}{\left| v_{x'} \right|+  \left| v_{y'} \right|} 
\label{v1c}
\end{equation} 
The decomposition thus specified is unique, and it ensures that the 
component $v_z$ is 
distributed along x'-z and y'-z planes proportionally to the 
corresponding projections of ${\bf v}$
along the $\hat{x}'$, and $\hat{y}'$ directions. It is easy to see
however, that any other decomposition of $v_z$  would lead to the
same results for the nonlinear currents derived below, except for
higher order corrections in $\delta^{-1}$ which shall be neglected in
any case.

The total phase space
contributing to the quasiparticle 
part of the current 
can be obtained by considering the  effects of ${\bf v}_1$ and ${\bf v}_2$ 
separately. Let us consider the effect of  ${\bf v}_1$ in  
(\ref{jqp}). Quasiparticle excitations are allowed in the region 
described by $\phi \leq \phi _c$, 
where $({\bf v}_f \cdot {\bf v}_1)^2= (\Delta_0 \sin \phi _c)^2$ and 
\begin{equation}
\phi_c^2= \frac{v_f^2 v_{x'}^2}{4 \Delta_0^2} (\sin ^2 \theta+ 
\frac{2}{ \sqrt{\delta}} \sin \theta \cos \theta \cot \omega 
\frac{\left| v_{x'} \right|} {v_{x'}}+\frac{ 1} {\delta} \cos ^2 
\theta \cot ^2 \omega)
\label{phic}
\end{equation}
where we have approximated  the order parameter 
in the nodal region by $\Delta(\phi) \approx \Delta _0 2 \phi$.
It would be easy to write instead $\Delta(\phi)\approx\Delta_0\mu \phi$,
with $\mu$ being a free parameter representing the slope of the OP
function near the node, as was done in Ref.\onlinecite{ys}. This can
be viewed as modifying the characteristic field $H_0$ (Eq. (\ref{h0})
by a factor of $\mu/2$.

The integrals involved in the calculation of the
nonlinear contribution to the relation 
${\bf j(v)}$ due to the ${\bf v}_1$ are then of the form
\begin{equation}
I_{x'}=2 \Delta_0 \int_0^{\pi} d \theta \sin^2 \theta \int_{-\phi_c}^{\phi_c} 
d \phi \: v_f  \sqrt{\phi_c^2 - \phi^2}  
\end {equation}
\begin{equation}
I_{z1}=2 \Delta_0 \int_0^{\pi} d \theta  \sin \theta \cos \theta
\frac{1} {\sqrt{\delta}}
 \int_{-\phi_c}^{\phi_c} d \phi \:
v_f  \sqrt{\phi_c^2 - \phi^2} 
\end {equation}
After integration over the angles $(\theta, \phi)$ we get:
\begin{equation}
 I_{x'}= \frac {3 \pi ^2}{32} v_f^2 \frac {v_{x'}^2}{v_c}
       (1+ \frac {\cot ^2 \omega}{3 \delta}) \label{ix}
\end{equation}
\begin{equation}
 I_{z1}= \frac {\pi ^2}{16 \: \delta} v_f^2 \frac {1}{v_c}
       (v_{x'}^2 \cot \omega) \label{iz1}
\end{equation}
The contribution to the quasiparicle current due 
to ${\bf v}_2$, is calculated in a precisely similar manner and
it gives analogous expressions for $I_{y'}$ and $I_{z2}$.
Combining the contributions due to  ${\bf v}_1$, 
${\bf v}_2$ and omitting, as indicated above, subleading  
terms of order $(1/ \delta)$ in Eq. (\ref{ix}) we get:
\begin{equation}
j_{qp \:x',y'}=e N_f v_f^2 \frac {3 \pi }{64}  \frac {v_{x',y'} 
\left|v_{x',y'}\right|}{v_c}
 \label{jqpx}
\end{equation}
\begin{equation}
 j_{qp \:z}= e N_f v_f^2 \frac { \pi }{32 \: \delta} \frac {1}{v_c}
       (v_{x'}^2 \cot \omega+v_{y'}^2 \cot  \omega) \label{jqpz}
\end{equation}
In the above expression we substitute $\cot \omega$ given in  
(\ref{v1c}) and  recover the nonlinear part of Eq. (\ref{jnl}).

If, as an alternative, we consider an OP function of the naive 3-d
form\cite{sigrist} $\propto (k_x^2-k_y^2)$:
\begin{equation}
\Delta(\theta,\phi)=\Delta_1 \sin^2(\theta)\sin(2\phi)
\label{opalt}
\end{equation}
then all integrals can still be done and one obtains instead:
\begin{equation}
j_{qp \:x',y'}=e N_f v_f^2 \frac {\pi }{16}  \frac {v_{x',y'} 
\left|v_{x',y'}\right|}{v_c}
 \label{jqpxa}
\end{equation}
\begin{equation}
 j_{qp \:z}= e N_f v_f^2 \frac { \pi }{8 \: \delta} \frac {1}{v_c}
       (v_{x'}^2 \cot \omega+v_{y'}^2 \cot  \omega) \label{jqpza}
\end{equation}
where $v_c$ is now defined in terms of $\Delta_1$.
The important  nonlinear coefficient in (\ref{jqpxa}) is a factor of $4/3$
larger than that in \ref{jqpx}. The results for ${\cal M}$ increase by the
same factor. However, in Eq. (\ref{h0}) the quantity $\xi_0$ should be
replaced by some average such as $(\xi_{ab}^2 \xi_c)^{1/3}$
which is smaller by about the same amount.
\section{Quantities in spheroidal coordinates }

The superfluid density tensor, given in the obvious
cartesian coordinates by a diagonal tensor with components 
$(\rho_{ab}, \rho_{ab}, \rho_c)$, is converted to oblate spheroidal
coordinates by  performing the appropriate transformation. One obtains:
  \[ \tilde{\rho} = \rho_{ab}
\left[ \begin{array}{ccc} \frac{\sinh^2\alpha \sin^2\beta +  \delta^{-1}
  \cosh^2\alpha \cos^2\beta}{\sinh^2\alpha+\cos^2\beta} &
\frac{(1-\delta^{-1})\sinh\alpha \cosh\alpha
\sin\beta \cos\beta}{\sinh^2\alpha+\cos^2\beta} & 0 \\
\frac{(1-\delta^{-1})\sinh\alpha \cosh\alpha
\sin\beta \cos\beta}{\sinh^2\alpha+\cos^2\beta} & 
\frac{\delta^{-1}\sinh^2\alpha \sin^2\beta + 
\cosh^2\alpha \cos^2\beta}{\sinh^2\alpha+\cos^2\beta} & 0 \\
0 & 0 & 1\end{array} \right] \]      

To solve equations (\ref{maxlon}) the expression 
$\nabla\times\nabla\times {\bf v}$ should be
transformed to oblate spheroidal coordinates:
\begin{eqnarray}
f^2 (\nabla\times\nabla\times {\bf v})_{\alpha} &= 
a_0\partial_{\beta\beta}v_{\alpha}+ 
a_1\partial_{\beta}v_{\alpha}+  
a_2\partial_{\phi\phi}v_{\alpha}+
a_3v_{\alpha}+ 
a_4\partial_{\alpha\beta}v_{\beta}+ 
a_5\partial_{\alpha}v_{\beta}+ 
a_6\partial_{\beta}v_{\beta} \nonumber \\ 
&+a_7v_{\beta}
+a_8\partial_{\alpha\phi}v_{\phi}+a_9\partial_{\phi}v_{\phi} 
\end{eqnarray}

\begin{eqnarray}
f^2 (\nabla\times\nabla\times {\bf v})_{\beta} &= 
b_0\partial_{\alpha\beta}v_{\alpha}+ 
b_1\partial_{\alpha}v_{\alpha}+ 
b_2\partial_{\beta}v_{\alpha}+b_3v_{\alpha}+ 
b_4\partial_{\alpha\alpha}v_{\beta}+ 
b_5\partial_{\alpha}v_{\beta}+  
b_6\partial_{\phi\phi}v_{\beta} \nonumber \\ 
&+b_7v_{\beta}+ 
b_8\partial_{\beta\phi}v_{\phi}+b_9\partial_{\phi}v_{\phi} 
\end{eqnarray}

\begin{eqnarray} 
f^2 (\nabla\times\nabla\times {\bf v})_{\phi} &= 
p_0\partial_{\alpha\phi}v_{\alpha}+ 
p_1\partial_{\phi}v_{\alpha}+ 
p_2\partial_{\beta\phi}v_{\beta}+ 
p_3\partial_{\phi}v_{\beta}+ 
p_4\partial_{\alpha\alpha}v_{\phi}+ 
p_5\partial_{\beta\beta}v_{\phi} \nonumber \\  
&+p_6\partial_{\alpha}v_{\phi}+ 
p_7\partial_{\beta}v_{\phi}+p_8 v_{\phi}
\end{eqnarray}  
we recall that f is a  focal length scale factor. 
The coefficients $a_i,b_i,p_i$ are given by (using the abreviations
$t\equiv\sinh\alpha$, $u\equiv\cosh\alpha$, 
$s\equiv\sin\beta$, $c\equiv\cos\beta$, $w\equiv(t^2+c^2)^{1/2}$): 
\begin{eqnarray}  
&a_0=-a_4=-b_0=b_4=p_4=p_5=-\frac {1}{w^2} \nonumber \\ 
&a_1=-\frac{w^2}{u^2}a_5=p_7=\frac{ws}{u}p_3=-\frac{c}{w^2 s} \nonumber \\ 
&a_2=b_6=-p_8=-\frac{1}{u^2 s^2} \nonumber \\
&a_3=\frac{2 c^2-t^2+3 t^2 c^2}{w^6} \nonumber \\
&a_6=\frac{t u }{w^4} \nonumber \\
&a_7=\frac{tuc(3+t^2-2c^2)}{w^6 s} \nonumber \\
&a_8=b_8=p_0=p_2=\frac{u}{t}a_9=\frac{s}{c}b_9=\frac{1}{u w s} \nonumber \\
&b_1=-\frac{c s}{w^4} \nonumber \\
&b_2=\frac{u s}{w}p_1=-\frac{t s^2}{u w^4}\nonumber \\
&b_3=\frac{tcs(3+2 t^2-c^2)}{w^6 u} \nonumber \\
&b_5=p_6=-\frac{2t}{u w^2} \nonumber \\
&b_7=\frac{2t^2-c^2-3 t^2 c^2}{w^6} \nonumber
\end{eqnarray}   
The right hand side of (\ref{maxlon}) has to be transformed to spheroidal
coordinates and added to the expressions above. 

%
%
\begin{figure}
\caption{Geometry 
considered in this paper. The sample is represented by an oblate
ellipsoid of revolution. The left half of this figure shows the top
view, and the right half a side view. The $x$, $y$ and $z$ directions
are fixed in space. The field is applied along the $x$ axis, as 
indicated, while ${\bf m_\perp}$ is along the $y$ axis. The sample is
rotated about the $z$ direction. The long and
short semiaxes values are called 
$A$ and $C$ in the text, respectively.}
\label{fig1}
\end{figure}
 \begin{figure}
\caption{The cartesian component
of the current, $j_z$, as a function of distance from the
surface, beginning a the point with cartesian coordinates $x=0$, $y=A$, $z=0$.
The quantity plotted is the current normalized to its value at the surface.
$D$ is distance in units of $\lambda_{ab}$: $D\equiv (A-y)/\lambda_{ab}$.
This current is dominated by linear effects in the field.
One can see that its decay is governed by $\lambda_c$, which is
$\lambda_c=4\lambda_{ab}$ for the example plotted here, where we have used
$H/H_0=0.1$ and $A/C=7$.}
 \label{linearcur}
 \end{figure}
\begin{figure}
\caption{The linear and nonlinear components of the current as a function
of distance from the surface beginning at the point $x=y=0$, $z=C$. The dotted
line is $j_y$ normalized to its value at the surface. It is predominantly
linear and its decay is governed by $\lambda_{ab}$. The solid line represents
the nonlinear component $j_x$ along the field direction, also normalized
to its own, much smaller,
surface value. Its nontrivial behavior is discussed in the text.
$D\equiv (C-z)/\lambda_{ab}$ and all other parameters are as in the
previous Figure.}
\label{nonlincur}
\end{figure}
\begin{figure}
\caption{Predictions of our theory for a typical HTSC single crystal
sample. The size, shape, and material parameters are indicated in the
text. The quantity plotted
 is the maximum value
 of $4 \pi m_\perp$, 
(in units of $10^{-6}$  gauss cm$^3$)
as the sample is rotated,
 as a function of applied field in gauss.}
\label{predict}
\end{figure}
\begin{figure}
\caption{The angular dependence of the transverse magnetization,
normalized to unity at its maximum. 
The points are our numerical results at
several angles. The solid
line represents the analytic result for the angular dependence in the 
purely two-dimensional case, normalized in the same way.
Within error bars the angular dependence is the same,
although the amplitude changes.}
\label{shape}
\end{figure}
\begin{figure}
\caption{Dependence of the transverse moment
amplitude  on the applied field. The quantity
plotted vs dimensionless field $H/H_0$,
is $G\equiv 10^2 4\pi m_\perp/(H\lambda_{ab}{\cal S})$
(see Eq. (\protect\ref{mtr2}).)
The symbols are our numerical results and the dashed line
the best linear fit. The linear dependence of $G$ on $H$
means that ${\cal M}$ is independent of field, and hence
that $m_\perp$ is quadratic in $H$.}
\label{hsquare}
\end{figure}

%
%
 \begin{table}
 \caption{The dimensionless quantity ${\cal M}$, defined by
Eq. (\protect{\ref{mtr2}}), as a function of the material parameter
$\delta$ and of sample shape.}
 \label{table1}
 \begin{tabular}{cccc}
$\delta=(\lambda_c/\lambda_{ab})^2$&A/C=19&A/C=10&A/C=7\\
\tableline
16&0.061&0.060&0.059\\
25&0.058&0.057&0.056\\
36&0.057&0.054&0.052\\
50&0.055&0.051&0.049\\
\end{tabular}
\end{table}


\begin{references}
\bibitem{agl} J. Annett, N. Goldenfeld, and A. J. Leggett, to appear in
{\it Physical properties of High Temperature Superconductors},
World Scientific, Singapore, (1996).
\bibitem{buanyang} J. Buan {\it et al} preprint.
\bibitem{ysprl}S.K.  Yip and J.A. Sauls, Phys. Rev. Lett. {\bf 69}, 2264,
(1992).
\bibitem{sv}B.P. Stojkovi\'c and O.T. Valls, Phys. Rev. {\bf B51}, 6049,
(1995).
\bibitem{ys}D. Xu, S.K. Yip and J.A. Sauls, Phys Rev. {\bf B51}, 16233,
(1995).
\bibitem{bahcall}See for example, S.R. Bahcall, unpublished.
\bibitem{confusion} Magnetic moment is magnetization times volume. As both
quantites are given in ``emu'' in gaussian units, much confusion between
them occurs in the literature. Here
we will give results in terms of 
$4\pi m$, in gauss times volume, as it is in fact done in the 
experimental work cited here. 
\bibitem{buan} J. Buan, B.P. Stojkovi\'c, N.E. Israeloff, C.C. Huang, A.M.
Goldman, and O.T. Valls, Phys. Rev. Lett. {\bf 72}, 2632, (1994).
\bibitem{orlando} T.P. Orlando and K.A. Delin, {\it Foundations of Applied
Superconductivity}, Addison-Wesley, Reading, (1991).
\bibitem{amgpc} A.M. Goldman, private communication.
\bibitem{lin}Z.H. Lin, G.C. Spalding, A.M. Goldman, B.F. Bayman, and
O.T. Valls,  Europhysics Lett. {\bf 32}, 573, (1995).
\bibitem{degennes} P.G. DeGennes, {\it Superconductivity of Metals and Alloys}
Addison-Wesley, reading, (1989).
\bibitem{reitz}J.R. Reitz, F.J. Milford, and R.W. Christy, {\it Foundations
of Electromagnetic Theory}, Addison-Wesley, Reading, (1989), page 329.
\bibitem{jackson} J. D. Jackson, {\it Classical Electrodynamics},
John Wiley, New York, (1975), page 181.
\bibitem{jackson2}J. D. Jackson, {\it op. cit.} inside cover.
\bibitem{fetwal} A. Fetter and J.D. Walecka, {\it Theory of Many Particle
Systems}, McGraw-Hill, New York, (1971).
\bibitem{london} F. London, {\it Superfluids}, v 1, John Wiley, New York,
(1950).
\bibitem{bardeen}J. Bardeen, Rev. Mod. Phys. {\bf 34}, 667, (1962).
\bibitem{kim} D.N. Basov {\it et al}, Phys. Rev. Lett. {\bf 74}, 598, (1995).
K. Zhang {\it et al}, Phys. Rev. Lett. {\bf 73}, 2484, (1994).
\bibitem{erange} See e.g. N.M. Plakida,
{\it High Temperature Superconductivity}, Springer, Berlin, (1995).
\bibitem{film} In the limit $\lambda\sim d$ the nonlinear effects disappear
very quickly, as the sample becomes essentially normal.
\bibitem{bgg} A. Bhattacharya, D. Grupp, A.M. Goldman,
and U. Welp, submitted to Applied Physics Letters.
\bibitem{arfken} G. Arfken, {\it Mathematical Methods for Physicists},
Academic Press, New York, (1970), Chapter 2. This subject was
dropped in later editions.
\bibitem{russian} N.N. Lebedev, I.P. Skalskaya, and Y.S. Uflyand,
{\it Worked Problems in Applied Mathematics}, Dover, New York, (1979).
\bibitem{ll} L.D. Landau and E.M. Lifshitz, {\it Electrodynamics of
Continuous Media}, Pergamon, Oxford, (1960).
\bibitem{flat} In the ``slab'' case this procedure would yield immediately
the correct result, since then there is no ``radial'' field and no
boundary condition is sacrificed.
\bibitem{zv} Additional mathematical details will be given elsewhere.
\bibitem{noecc} One could use the eccentricity instead, but $C/A$ is more
immediately accessible.
\bibitem{scaling} The computational systems
used, on which this scaling has been verified, range in size all the way up to
the actual size of the physical samples.
\bibitem{sigrist} M. Sigrist and T.M. Rice, Zeitschrift fur Physik B 
(Condensed Matter) {\bf 68}, 9, (1987)
\bibitem{anandpc} A. Bhattacharya private communication.
\bibitem{muz} C.H. Choi and P. Muzikar, Phys. Rev. {\bf B39}, 11296, (1989).
\end{references}
\end{document}